# Educational impacts of generative artificial intelligence on learning and performance of engineering students in China

Lei Fan*, Kunyang Deng and Fangxue Liu

Department of Civil Engineering, Design School, Xi'an Jiaotong-Liverpool University, Suzhou, China

*Corresponding Author (Email: Lei.Fan@xjtlu.edu.cn)

**Abstract:** With the rapid advancement of generative artificial intelligence (AI), its potential applications in higher education have attracted significant attention. This study investigated how 148 students from diverse engineering disciplines and regions across China used generative AI, focusing on its impact on their learning experience and the opportunities and challenges it poses in engineering education. Based on the surveyed data, we explored four key areas: the frequency and application scenarios of AI use among engineering students, its impact on students' learning and performance, commonly encountered challenges in using generative AI, and future prospects for its adoption in engineering education. The results showed that more than half of the participants reported a positive impact of generative AI on their learning efficiency, initiative, and creativity, with nearly half believing it also enhanced their independent thinking. However, despite acknowledging improved study efficiency, many felt their actual academic performance remained largely unchanged and expressed concerns about the accuracy and domain-specific reliability of generative AI. Our findings provide a first-hand insight into the current benefits and challenges generative AI brings to students, particularly Chinese engineering students, while offering several recommendations—especially from the students' perspective—for effectively integrating generative AI into engineering education.

**Key words**: artificial intelligence; pedagogy; learning; student; engineering

## 1. Introduction

Generative artificial intelligence (AI), such as ChatGPT developed by OpenAI, has gained significant attention for its innovative capabilities. By leveraging deep learning, generative AI creates diverse content, including text, images, audio, and video, excelling in creative and interactive tasks [1]. In education, it offers immense potential for personalized learning, instant feedback, and assistance in tasks like data analysis, literature review, and report writing, thereby enhancing learning outcomes [2, 3, 4, 5]. Its ability to support complex problem-solving and research development through precise outputs and simple instructions makes it a transformative tool, particularly in engineering education [6, 7, 8].

Engineering education is founded on an integration of multiple disciplines such as engineering, science, technology, and mathematics [9, 10]. It connects theoretical concepts to real-world applications [11]. Traditional engineering education often emphasizes theory-based teaching and ignores students' initiative in the learning process, which falls short in helping students develop practical problem-solving skills. Self-determination theory states that learners' intrinsic motivation will be enhanced by satisfying the three psychological needs of autonomy, competence, and relatedness, thereby improving learning engagement and effectiveness [12,13]. Generative AI significantly enhances engineering education by addressing gaps in traditional theory-driven teaching and meeting these three psychological needs through adaptive learning that provides personalized learning paths.

While generative AI holds great promise, its integration into education also raises critical ethical and pedagogical concerns. Issues such as plagiarism, academic integrity, and fairness in assessment have become increasingly prominent [14, 15, 16]. It is widely recognized that generative AI tools have been misused to produce unoriginal content, raising questions about how to detect and prevent such practices effectively. Although tools like GPT Zero and Originality AI have been developed to identify AI-generated text, their accuracy remains inconsistent [16].

Although generative AI is capable of learning from substantial amounts of data, their training data might contain inherent biases, inaccurate content, or outdated information that can be replicated in the generated content. Consequently, the output generated by generative AI tools may not always be accurate, especially in engineering disciplines that are highly technical and demand a high degree of accuracy, and the incorrect output might lead to misleading comprehension for students. Therefore, how to ensure that students possess adequate judgment when employing AI tools and avoid blindly accepting AI-generated content has emerged as an urgent problem to be addressed [17, 18, 19]. In addition, students' over-reliance on AI tools may lead to a deterioration in independent thinking, which contradicts the core objective of engineering education to cultivate critical thinking [9]. Therefore, how to effectively incorporate generative AI in educational practice while avoiding its potential negative effects is a question worthy of in-depth discussion.

Effectively addressing these issues is essential for defining the role of generative AI in engineering curricula in higher education. Institutions and educators must develop comprehensive guidelines and encourage responsible AI use, guiding students to utilize AI tools effectively and ethically while fostering innovative learning and teaching models [3, 20]. Achieving this goal requires educators to adopt informed strategies grounded in a thorough understanding of how engineering students currently engage with AI.

Although previous studies have explored the role of generative AI in teaching [21,22], there is still a lack of systematic analysis on how it affects students' learning initiative, critical thinking, and self-cognition, especially in the field of engineering education that focuses on practical applications. Furthermore, technology adoption is not only a choice at the tool level, but also a reflection of the interaction between learning motivation, educational culture, and social norms [23,24,25]. In this regard, how engineering students use generative AI and how generative AI is embedded in their daily learning process needs to be understood in the context of the specific Chinese social culture and engineering education system. Therefore, the significance of this study is to fill the current empirical research gap on the impact of generative AI on Chinese students' learning behavior and motivation in engineering education. While this study does not aim to propose new theoretical concepts, its valuable data and observations contribute to a deeper understanding of established pedagogical theories and may also inform the development of new theories in this field.

By analyzing questionnaire data from diverse academic backgrounds, the study evaluates students' perceptions of AI use, highlights key challenges currently encountered by them, and proposes effective integration strategies for engineering curricula. The scope of the study covers students' frequency of use of AI, purpose of use, perceived impact on their learning behaviors and outcomes, and commonly encountered challenges. Specifically, it addresses the following research questions:

(1) How frequent and in what ways are engineering students in China utilizing generative AI?
(2) What is the impact of generative AI on the learning experience of Chinese engineering students?
(3) What challenges do Chinese engineering students currently face when using generative AI, and what are their expectations for its integration into engineering education?

The rest of this article is organized as follows: Section 2 provides a brief overview of current application of generative AI and related educational theories. Section 3 details the design and sampling strategy of the questionnaire survey. Section 4 presents the reliability and validity analysis of the collected survey data. Section 5 quantitatively presents and discusses the survey results. Section 6 discusses the theoretical and practical implications of this study, outlines the study's limitations, and proposes future research directions. Section 7 concludes the study.

## 2. Literature review

With the advancement of AI technologies, generative AI has attracted significant attention across a range of academic and professional disciplines [26]. Powered by models such as the Transformer architecture, generative AI can automatically produce diverse forms of content, including text, images, and audio, based on input data, and is capable of simulating human-like reasoning and expression to a considerable extent [27]. Its integration into real-world applications is already evident across numerous industries. For instance, generative AI supports personalized learning in education [4, 5], diagnostic assistance in healthcare [28], environmental pollution assessment in environmental engineering [29], and infrastructure damage evaluation in the construction sector [30, 31].

In the education sector, the ability of generative AI to deliver personalized and adaptive learning experiences gives it unique advantages. It enables a broad array of learning opportunities such as real-time feedback, automated teaching resource generation, adaptive content delivery, and interactive learning environments, all of which contribute to a more engaging and customized educational experience [9, 32, 33]. For example, Museanu (2024) [34] reported that AI-driven translation tools and interactive platforms significantly enhanced student engagement and accelerated foreign language acquisition. Similarly, an exploratory study by Sun and Deng [35] involving 70 U.S. business students found that ChatGPT improved task efficiency and user satisfaction. However, its effectiveness was influenced by students' prior experience and the design of instructional prompts.

Generative AI also facilitates problem-based and collaborative learning through hands-on activities that encourage students to actively apply theoretical knowledge in practical contexts [36]. By supporting the development of critical thinking, reflective communication, and problem-solving skills, generative AI helps prepare students for complex future challenges [32, 33]. Furthermore, it strengthens students' understanding of abstract engineering concepts and boosts engagement in experimental and design-based projects by promoting active and reflective learning practices [37, 38, 39].

As generative AI continues to expand in educational settings, feedback mechanisms have emerged as a critical area of research. The value of immediate feedback was recognized as early as 1981 when psychologist Skinner [40] proposed reinforcement learning theory, which emphasizes that learning efficiency is greatly enhanced by timely and targeted feedback. He advocated for structured instructional designs that provide feedback at each step of the learning process. However, delivering real-time feedback to large groups of students remains a challenge for educators. In response, recent advances in automated feedback systems have been promising. For example, Escalante et al. [41] applied generative AI to offer multi-dimensional suggestions in automated writing evaluation systems, while Hao et al. [42] developed AI-based grading tools. Nonetheless, these systems often fall short in providing deep cognitive insights, particularly in helping students identify and address core issues in argumentative writing [43, 44]. A potential path forward lies in blending human expertise with generative AI feedback to enhance the quality and depth of student support.

In parallel, various educational theories have been used to understand and enhance AI's role in learning. Constructivist learning theory, for example, views learning as an active process of knowledge construction through exploration, reflection, and critical inquiry [45, 46]. Based on this theory, Winkler et al. [47] examined how constructivist approaches support students in solving problems independently. Banihashem et al. [48] extended this work by designing an instructional model that integrates constructivist principles with learning analytics, effectively addressing issues related to low student engagement and poor self-regulated learning.

Self-determination theory [12, 13] is especially relevant in engineering education, as it emphasizes the importance of fulfilling learners' psychological needs for autonomy, competence, and relatedness. These needs can be supported by generative AI through adaptive, personalized, and interactive learning environments. In addition, frameworks such as the technology acceptance model (TAM) [49, 50] and multiple intelligences theory [51] have also been widely applied to evaluate students' attitudes toward AI tools and their perceived effectiveness in diverse educational contexts.

## 3. Methodology

### 3.1 Questionnaire Design

This study employs an anonymized questionnaire survey ($n = 148$) to examine the use of generative AI among Chinese engineering students, evaluate their perceptions, summarize their encountered challenges and explore its potential future integration into engineering curricula in higher education. The quantitative survey data provide a broad understanding of usage patterns on generative AI's role in engineering education in China.

To ensure a diverse sample, the study includes participants from multiple geographical regions within China, representing various institution types (e.g., double first-class universities, general universities, vocational colleges), engineering disciplines (e.g., computer science related fields, civil engineering, mechanical engineering, electrical engineering etc.), and educational levels (undergraduate and postgraduate).

The questionnaire mainly consists of 21 scale-based questions (provided in Appendix A) for precise quantification, 7 multiple-choice questions for specific information collection, and one open-ended question to allow respondents to share insights not anticipated in closed questions. These exclude questions related to participant's basic details such as gender, age, location, institution type, and major.

The scale-based questions are grouped into five variables:

(1) Frequency of generative AI use across scenarios
(2) Preference of using generative AI in complex problem-solving
(3) Impact of generative AI on individual learning capabilities
(4) Challenges and concerns about generative AI
(5) Perspectives on future applications of generative AI

### 3.2 Sampling strategy

The experiment protocol of this study was approved by the University Research Ethics Review Panel of Xi'an Jiaotong-Liverpool University on 21 September 2024. The survey was performed in accordance with relevant guidelines and regulations. Informed consent was obtained from all participants.

The study employed a combination of purposeful sampling and convenience sampling. Purposeful sampling is used to ensure representation from different institution types and different engineering disciplines (e.g., mechanical engineering, electrical engineering, computer science). Convenience sampling is used to obtain a broad range of responses from an easily accessible student population. Our sample consisted of 148 engineering students (n=148). The data were collected between October 9, 2024, and November 1, 2024, through both offline and online participation, with informed consent obtained from all participants. Each participant agreed to participate in the study knowing the purpose of the experiment and how their personal information would be used. The sampling strategy employed in this study is well-suited and adequate for its main objective of exploring the overall impact of generative AI on the learning behaviors and academic performance of Chinese engineering students. A discussion on alternative sampling approaches is also discussed in Section 6.3.

As shown in Table 1, the survey included 72 undergraduates and 76 postgraduates from diverse engineering disciplines in various geographical representations in China. The samples included 94 male (63.51%) and 54 female (36.49%) participants.

## 4. Reliability and validity analysis of surveyed data

Internal consistency reliability measures the degree of correlation among items within a questionnaire, typically assessed using Cronbach's α coefficient [52], which ranges from 0 to 1. A higher α value signifies stronger correlations between items and better internal consistency. $α > 0.8$ indicates excellent internal consistency (items are highly correlated and measure the same underlying construct). $0.7 ≤ α ≤ 0.8$ suggests good internal consistency, which is reliable for research and application. $0.6 ≤ α < 0.7$ infers acceptable reliability, which is usable but may require refinement. $α < 0.6$ highlights poor internal consistency, which needs revision.

In this study, 21 scale-based questions, within the five categorized variables, were analyzed for reliability. Cronbach's α was calculated for each of the five variables in the SPSS® Statistics software. The calculated Cronbach's α values for all variables exceeded 0.8 with an overall value of 0.879, confirming strong internal consistency across variables.

**Table 1.** Basic information on 148 participates, including their geographical distributions, engineering disciplines, gender and educational levels

| Category | Subcategory | Count/Percentage |
|---|---|---|
| **Geographical distribution** | North/Northeast China | 26.4% |
| | Central China | 39.1% |
| | Southwest/Northwest China | 14.8% |
| | South/Southeast China | 19.7% |
| **Engineering disciplines** | Computer Related Fields | 32.4% |
| | Civil Engineering | 17.6% |
| | Mechanical Engineering | 16.9% |
| | Electrical Engineering | 6.8% |
| | Other Disciplines (e.g., Chemical Engineering, Bioengineering) | 26.3% |
| **Gender distribution** | Male | 94 (63.5%) |
| | Female | 54 (36.5%) |
| **Educational levels** | Undergraduate | 72 (48.6%) |
| | Postgraduate | 76 (51.4%) |

**Table 2**: KMO for measuring sampling adequacy and Bartlett's test of sphericity

| **KMO value** | | **0.867** |
|---|---|---|
| **Bartlett's test of sphericity** | Chi-square $\chi^2$ | 1283.052 |
| | Degree of freedom *Df* | 210 |
| | *p* value | < 0.001 |

Following the reliability analysis, the validity of the questionnaire was assessed through the Kaiser-Meyer-Olkin (KMO) test [53] and the Bartlett's test [54]. The KMO test measures the adequacy of sampling for factor analysis. KMO values of 0.9 or above are considered very suitable, 0.8 to 0.9 are good, 0.7 to 0.8 are acceptable, and below 0.6 indicate poor suitability for factor analysis. Bartlett's tests test the null hypothesis that the correlation matrix is an identity matrix (no correlation among variables). If the *p*-value is less than 0.05, the null hypothesis is rejected, indicating that factor analysis can be performed.

The results of these two tests are shown in Table 2. The KMO value of 0.867 indicates good suitability for factor analysis. In the Bartlett's test, the *p*-value is 0.000, which is significant ($p < 0.05$), rejecting the null hypothesis and confirming that factor analysis is appropriate. These results demonstrate that the questionnaire data exhibit strong validity, enabling factor analysis with confidence.

## 5. Result

### 5.1 Frequency and purpose of using generative AI by Chinese engineering students

#### 5.1.1 Popularity of adopted generative AI tools

A significant portion of respondents reported first encountering generative AI within 1-2 years (i.e., 2023-2024). This reflects the rapid advancements in AI technology and its growing integration into educational and social contexts over the past couple of years in China. As the technology matures and its applications expand, more students have become aware of its potential and have started using it.

As shown in Figure 1, among various generative AI tools, ChatGPT was by far the most popular, with 77.03% of respondents using it. Its widespread adoption can be attributed to its powerful capabilities, user-friendly interface, and effectiveness in text generation and interaction. Following ChatGPT, Baidu's Wenxin Yiyan, which was particularly favored by Chinese engineering students, was used by 41.89% of respondents, likely due to its cultural and language advantages. Tools like DeepL (25.68%), Microsoft Bing (20.27%) and Google Bard (18.92%) also maintained a notable presence, especially for tasks like translation and search optimization. In contrast, tools such as DALL·E, Canva AI, Adobe Firefly, and others were less commonly used, each having usage rates around or below 10%, suggesting that their functions are not as aligned with the needs of engineering students in China.

It is worthwhile acknowledging that new generative AI tools have been rapidly emerging since the survey was completed in November 2024. A prominent example is DeepSeek, which has gained significant popularity, particularly in China. This development is likely to influence the AI tool preferences of Chinese engineering students. Nevertheless, the findings of this study remain valid, as the capabilities of these newly introduced tools are largely comparable to those of existing ones.

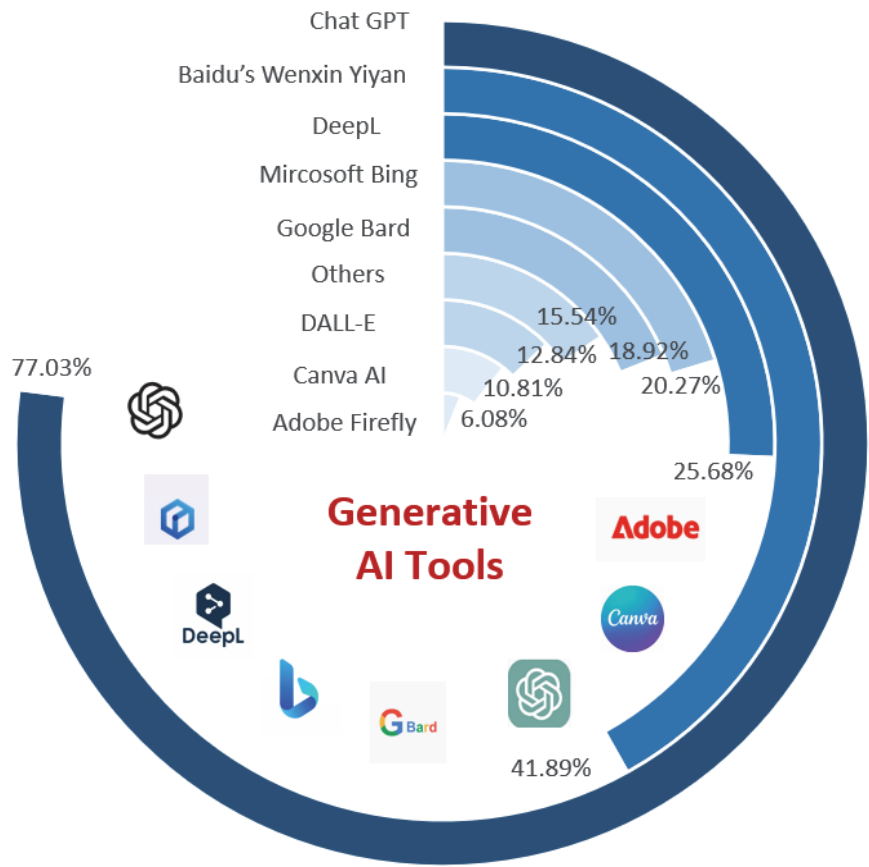

**Figure 1**. Popularity of generative AI tools among engineering students in China as of November 2024

**5.1.2 Frequency of using generative AI**

Generative AI has become a frequent tool in the academic routines of many engineering students in China, with 20.95% using it daily or 41.89% using it multiple times per week, as illustrated in Figure 2(a). Only 4.73% of respondents reported never using generative AI. Figure 2(c) suggests that the frequency of use was higher among postgraduate students (40% reporting regular use) compared to undergraduates (only 25.64%), likely due to the greater complexity of postgraduate-level tasks such as research, which require more advanced tool support. Additionally, our survey results also reveal that students in computer-related disciplines used generative AI more frequently, reflecting their stronger technical alignment with the tools and a deeper understanding of AI technologies.

**5.1.3 Application scenarios of generative AI**

Figure 2(d) illustrate that generative AI has been widely used in various academic tasks by Chinese engineering students. The most common application was "finding learning resources or concept explanations" (55.41%), indicating that many students have turned to AI for help with complex concepts or course materials. Other popular uses include "compiling reports or documents" (54.73%) and "data analysis" (52.03%), reflecting the tool's role in enhancing academic writing and data processing. The frequency of use stated in Section 5.1.2 and the application scenarios suggest the overwhelming majority of respondents have used generative AI for tasks related to their majors, such as solving engineering problems, optimizing design plans, or writing research reports.

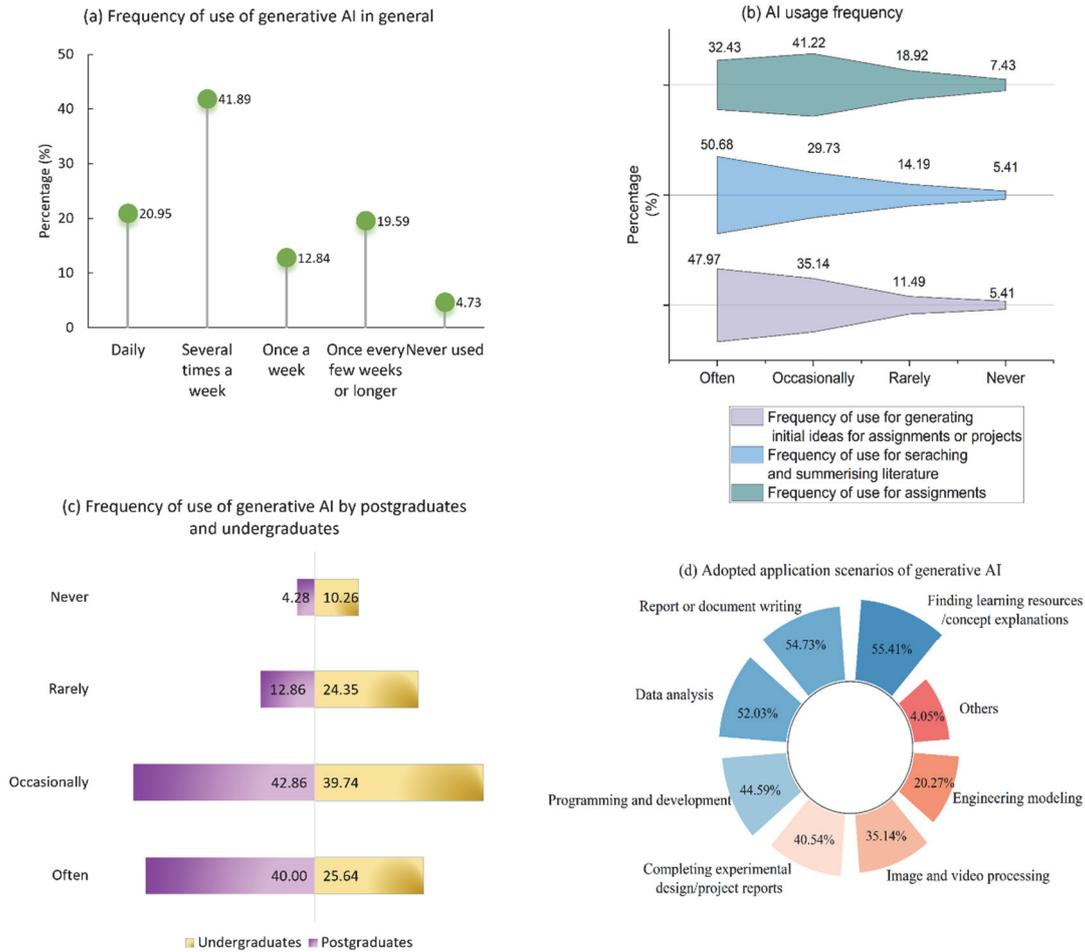

**Figure 2**. Current utilization status of generative AI tools by Chinese engineering students: (a) Frequency of use for general purposes, (b) Frequency of use for basic academic tasks, (c) Frequency of general use by postgraduates and undergraduates, (d) Typical adopted application scenarios.

When it comes to using generative AI for assignments, 32.43% of students used it frequently, and 41.22% used it occasionally, as shown in Figure 2(b). This indicates that generative AI has become a regular part of their academic workflows. Despite most surveyed students viewed AI as an important aid in their academic work, a minority (18.92% rarely and 7.43% never) were less inclined to use it, possibly due to doubts about its effectiveness or ethical concerns.

In terms of basic application scenarios of generative AI for engineering disciplines, as illustrated in Figure 2(b), 50.68% of respondents have often adopted AI tools for searching literature and assisting literature review, while 47.97% of the respondents turned to generative AI for generating initial ideas for their design assignment or research projects. These highlight the tool's role in supporting both academic work and creative thinking.

### 5.2 The impact of generative AI on Chinese engineering students' learning

#### 5.2.1 Impact on learning efficiency

As shown in Figure 3(a), the survey data reveals that 36.49% and 52.03% reported "significant improvement" and "improved", respectively, totaling 88.52% of respondents felt their learning efficiency had been improved by generative AI tools. Only 10.81% noted "almost no change," and a very small fraction (0.68%) felt that AI use had "reduced" or "significantly reduced" efficiency. These results indicate a strong consensus on the positive impact of AI on learning efficiency. These results provide additional support for previous studies (e.g., [40], [55]) highlighting that immediate feedback enables learners to recognize gaps in their understanding and proactively adapt their strategies to enhance learning efficiency. The rapid content generation, instant feedback, and extensive knowledge support offered by AI tools significantly reduced the time traditionally spent on learning and writing tasks, making them particularly valuable for time-sensitive academic activities like literature search, report writing and data analysis.

**5.2.2 Impact on active learning**

The questionnaire results show that generative AI tools had a significant positive impact on the active learning of Chinese engineering students. 64.19% of participants reported improved learning initiative, with 41.22% stating it had "improved" and 22.97% saying it had "significantly improved", as illustrated in Figure 3(a). This highlights that most students viewed generative AI as a valuable tool to boost motivation and engagement. However, 6.76% reported a decline in learning initiative, suggesting that AI could be detrimental for students with certain learning habits, such as over-reliance on the tool that may weaken critical thinking and active learning.

According to self-determination theory [12], frequent use of AI to complete learning tasks can undermine students' sense of autonomy and competence, weakening their intrinsic motivation to learn. This effect is further supported by the Overjustification Effect [56], which suggests that when intrinsically motivated activities are replaced by heavy reliance on external tools, individuals may lose interest in the activity itself. In this context, students who were once engaged in learning may begin to view generative AI merely as a means to quickly finish tasks, shifting their focus from genuine learning to mere task completion, and diminishing their interest and initiative over time. This emphasizes the need for caution when integrating generative AI into inclusive education. Clear usage guidelines should be established, positioning AI as a learning aid rather than a sole crutch, and educators should monitor its impact, especially on students with low engagement.

**5.2.3 Impact on independent thinking**

While generative AI tools have been widely regarded as efficiency boosters for learning, their impact on independent thinking shows a more complex pattern, as shown in Figure 3(a). About 34.46% of respondents felt that AI had "almost no change" on their ability to think independently, while 47.97% reported an "improvement" (23.65%) or "significant improvement (24.32)". This suggests that nearly half of the students viewed generative AI as a tool that enhances independent thinking by providing new perspectives, feedback, and access to additional knowledge. In addition, students typically needed to judge the accuracy of the AI generated content, which might further cultivate their independent thinking. However, 14.86% and 2.70% of respondents felt that AI had weakened or significantly weakened their ability to think independently. This reflects the potential downsides of over-reliance on AI, where students might bypass deep engagement with problems, opting instead for AI-generated solutions. Our findings further reinforce existing research, especially the studies by Zhai et al. [57] and Klingbeil et al. [58], which showed that when the reliability of generative AI advice is hard to assess, individuals are more inclined to trust it uncritically to avoid cognitive effort. In complex fields like engineering, this may result in a reduction of students' problem-solving skills and a lack of understanding of the underlying concepts.

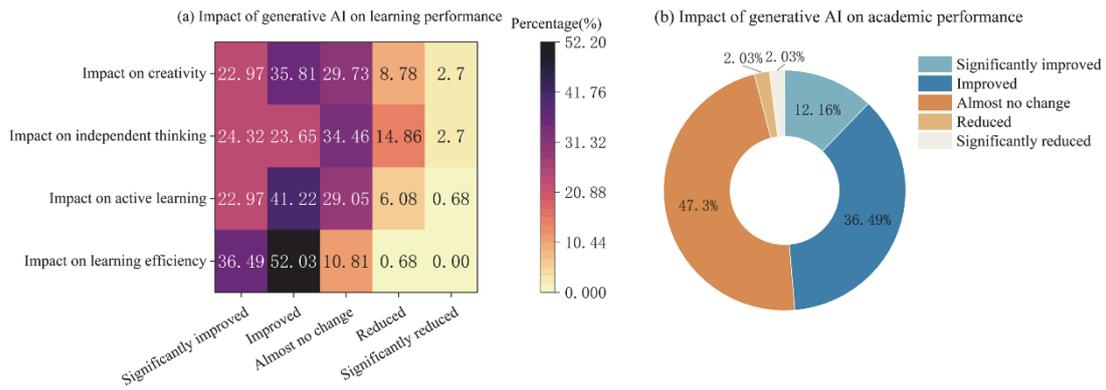

**Figure 3**. Impact of generative AI on Chinese engineering students' learning: (a) Impact on learning efficiency, initiative, independent thinking and creativity, (b) Impact on academic performance

### 5.2.4 Impact on creativity

When examining the impact of generative AI on creativity, the survey data show a similar pattern like AI's impact on independent thinking, as depicted in Figure 3(a). While 58.78% of respondents felt that AI had a positive effect on their creativity, with 35.81% noting an "improvement" and 22.97% a "significant improvement", a sizable portion (29.73%) reported "almost no change". This group likely used AI for more practical tasks, such as executing predefined solutions, rather than for creative inspiration. On the other hand, 11.48% of respondents believed AI had a negative impact on creativity, with 8.78% reporting a "reduction" and 2.7% a "significant reduction." These results suggest that while generative AI is widely seen as a powerful tool for generating ideas and solving complex problems, its ability to foster genuine innovation may be limited for some users, particularly if the generated content is too formulaic or restricts creative freedom. This can be explained through the theory of multiple intelligences [51]. Generative AI can mobilize multiple forms of intelligence, such as natural language understanding, human-like reasoning, and visual content, thereby improving one's creativity. However, differences in the intelligence structure of different individuals may lead to divergent effects of technology on creativity. However, differences in individuals' intelligence areas may result in varying effects of technology on creativity. If generative AI does not stimulate the intelligence areas that users rely on, they may feel that their creativity is limited.

### 5.2.5 Impact on academic performance

From Figure 3(b), It is notable that nearly half of the respondents did not feel that using generative AI had improved their academic performance, even though most reported enhanced learning efficiency and more active learning. This discrepancy can likely be attributed to the limited specialization of generative AI for specific engineering disciplines and concerns over the accuracy of generated content, as discussed in Section 5.3. It's important to recognize that perception doesn't always align with actual outcomes [59]. While generative AI often helps students complete tasks more quickly and smoothly, this can lead to a sense of "higher efficiency" by students. However, task completion doesn't necessarily indicate true understanding, and the satisfaction of finishing a task can mask the real extent of genuine knowledge learning. Over time, this immediate sense of accomplishment may create the illusion of "I have learned it," which can diminish students' continuous engagement in the learning process. These factors necessitate students' academic judgment in evaluating the relevance and accuracy of AI-generated materials. When paired with the strengths of generative AI, such judgment has significant potential to enhance academic performance, as reflected by 12.2% and 36.5% of respondents who reported "improved" or "slightly improved" academic outcomes, respectively.

**5.3 Challenges faced by Chinese engineering students in using generative AI**

**5.3.1 Key challenges in using generative AI**

Generative AI has significantly impacted learning efficiency and creativity, but its use in education presents several challenges. Key issues identified by engineering students include the accuracy of AI-generated content, over-reliance on AI tools, and technical difficulties, as shown in Figure 4.

The most prominent challenge, reported by 62.16% of respondents, was the inaccuracy of generated content. This lack of accuracy undermined students' confidence in using AI outputs, particularly for tasks involving complex data analysis. Despite AI's potential, there was still considerable room for improvement in content accuracy, especially tailored to specific engineering disciplines. This finding is consistent with existing theories suggesting that frequent exposure to inaccurate outputs, such as those from generative AI tools, can lead to "negative reinforcement," ultimately decreasing students' willingness to use these tools [40]. Moreover, the inaccuracy of automated feedback systems limits their ability to effectively support students in addressing more complex issues, such as developing critical thinking skills [43, 44]. As shown in Figure 5(a), when asked about the frequency of encountering inaccurate AI-generated content, 43.24% of respondents reported facing this issue often or very often, while 37% experienced it occasionally, highlighting the gap between students' needs and AI-generated content.

The second major concern, cited by 39.86% of respondents, was over-reliance on AI tools. Many students worried that excessive dependence on these tools might reduce their ability to solve problems independently and effectively.

Furthermore, 20.27% of respondents reported difficulties with the usability of AI tools, noting that the user interface can be difficult for beginners and that there was insufficient technical support, especially for the newly emerged AI tools. As indicated by the pedagogical theory TAM [49], when users, especially beginners, struggle with system navigation or functions, they are more likely to experience operational frustration, leading to reduced behavioral intention. Ethical concerns and privacy issues were raised by 14.19% of respondents, indicating continued unease about data protection and academic integrity. Additionally, 17.57% of students mentioned high costs as a barrier to accessing AI tools, which could limit their widespread use.

While generative AI offers clear advantages in education, the need for improvements in accuracy, usability, and cost is evident. These findings highlight areas for future development, with tool developers and educational institutions needing to focus on enhancing technical reliability, simplifying user interfaces, improving support systems, and reducing costs to make AI tools more accessible.

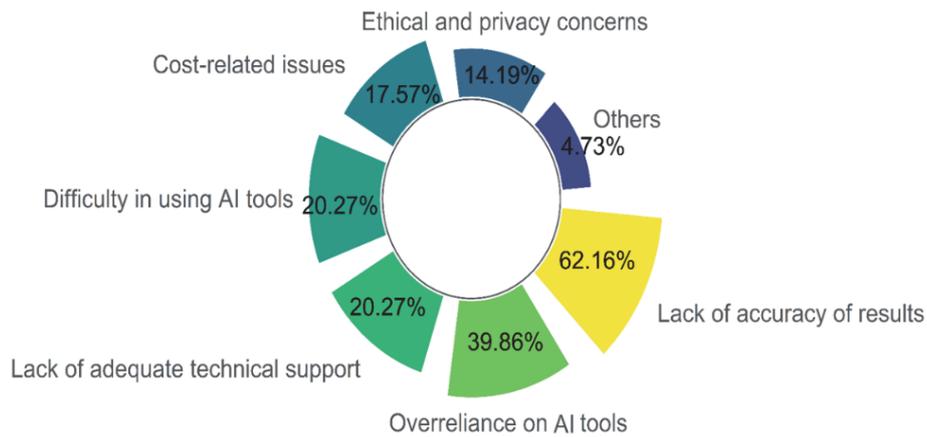

**Figure 4**. Key challenges in using generative AI by engineering students in China

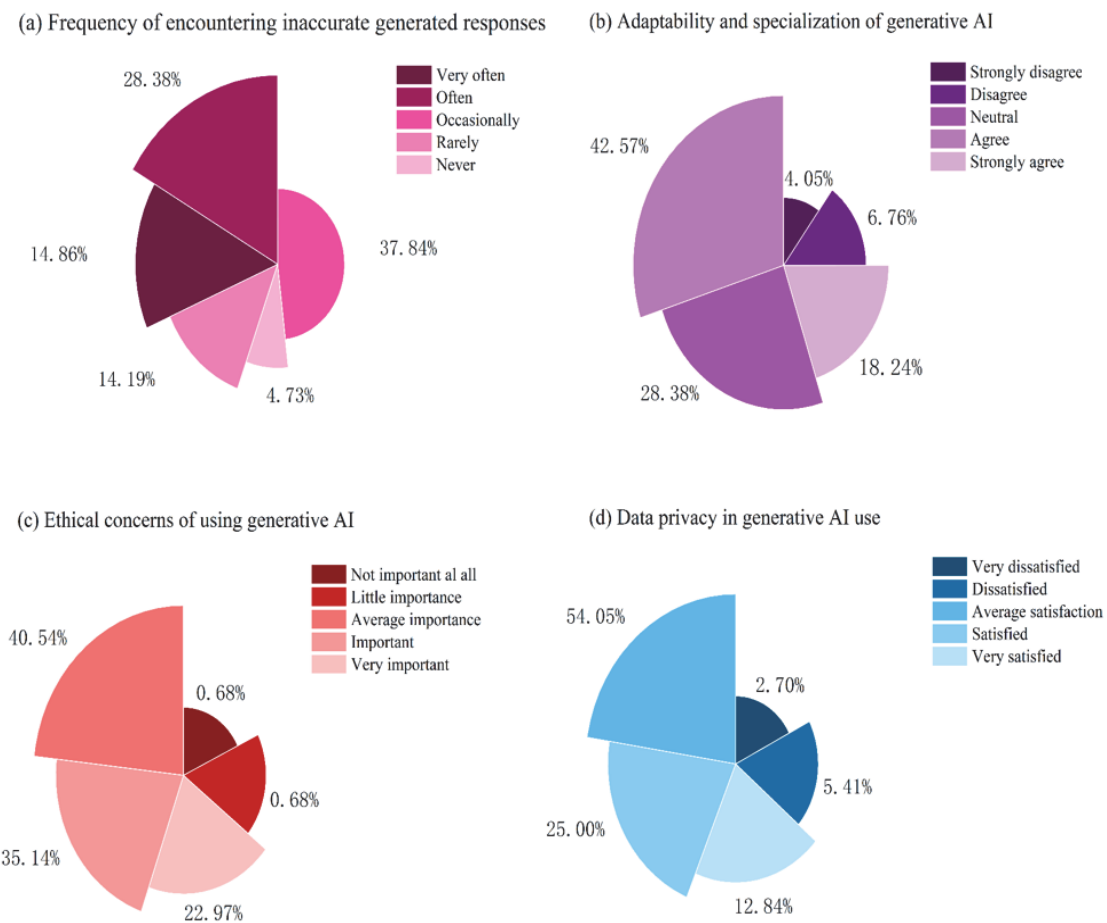

**Figure 5**. Specific concerns faced by Chinese engineering students: (a) Frequency of encountering inaccurate generated responses, (b) Adaptability and specialization of generative AI for meeting students' needs, (c) Ethics of using generative AI, (d) Data privacy associated with generative AI

### 5.3.2 Adaptability and specialization in engineering education

Despite rapid advancements in generative AI, its application in engineering education is constrained by the discipline's complexity and specialization. Mixed responses were received in our survey, as shown in Figure 5(b). The survey results showed that 42.57% of respondents "agree" generative AI tools were suitable for their professional needs, while 18.24% finding them "strongly agree" on their adopted AI tools.

However, nearly 40% of the respondents expressed concerns about the tools' effectiveness in addressing highly specialized problems. Specifically, 28.38% were "neutral" about AI's adaptability, and 16.81% deemed them "disagree" or "strongly disagree", suggesting concerns about AI's specialization towards specific engineering disciplines. These findings are consistent with prior research showing that when generative AI tools fail to recognize the knowledge structure, contextual logic, or deep conceptual relationships within a professional domain, they struggle to effectively support students in constructing meaningful knowledge [43, 44].

### 5.3.3 Ethical concerns and data privacy in generative AI use

Ethical issues surrounding the use of generative AI, especially in academic assignments, were another significant concern. Figure 5(c) revealed that 35.14% and 22.97% of respondents considered ethical issues "important" or "very important", respectively. However, 40.54% of respondents rated ethical issues as "average importance" and only 1.36% dismissed the importance of ethics. These findings suggest that ethical concerns regarding generative AI in education are widely acknowledged and should be addressed through regulation and guidance. Meanwhile, training on ethics is required for a broader recognition of various ethical issues.

Regarding data privacy, as shown in Figure 5(d), most respondents rated AI tools' performance as "average satisfaction" (54.05%), suggesting their lack of confidence in AI tools' privacy protection. In addition, 5.41% were "dissatisfied" and 2.7% were "very dissatisfied" with the data privacy, reflecting room for improvement in data privacy protection. Only 25% of the respondents found them "satisfied" and 12.84% "very satisfied" with data privacy, comparatively lower than their agreement on AI's effectiveness in aiding their learning. Concerns about privacy violations or ethical issues can discourage users from engaging with generative AI, even if the technology is perceived as useful [39]. These findings highlight the urgent need to provide relevant training to students on secure practices when using generative AI.

### 5.4 Expectation on generative AI use from Chinese engineering students

### 5.4.1 Attitudes towards generative AI integration into engineering education

Regarding the integration of generative AI into engineering education, as shown in Figure 6(a), 20.95% supported full integration into teaching, while 43.92% favored a partial integration that applies only to certain courses or scenarios. Additionally, 30.41% saw AI as a helpful tool, but not a necessity. Survey results show strong support for integrating generative AI into education. A large majority of respondents believed that generative AI courses should be compulsory for engineering students, indicating widespread recognition of its importance and value in engineering education.

Figure 6(b) shows mixed opinions on whether generative AI can replace traditional teaching models. 41.89% of respondents were "uncertain", while 12.84% thought it would "probably not" replace traditional methods. However, 12.84% believed that generative AI will eventually replace conventional teaching methods, while 30.41% supported this trend with some reservation. It appears that the overall support for complete replacement was not strong, likely due to technical limitations and ethical concerns surrounding generative AI. This finding indicates that the development and adoption of technology are not isolated processes but are deeply embedded within specific social contexts and norms. As previous studies [23, 24, 25] have reported, technology adoption is influenced by a range of factors, including learning motivation, educational culture, and prevailing social norms.

In terms of AI-related training offerings, Figure 7(a) 43.24% and 49.32% of respondents expected institutions to provide basic introductory AI courses and in-depth practical training on the use of generative AI, respectively. Nearly half of respondents preferred online training courses.

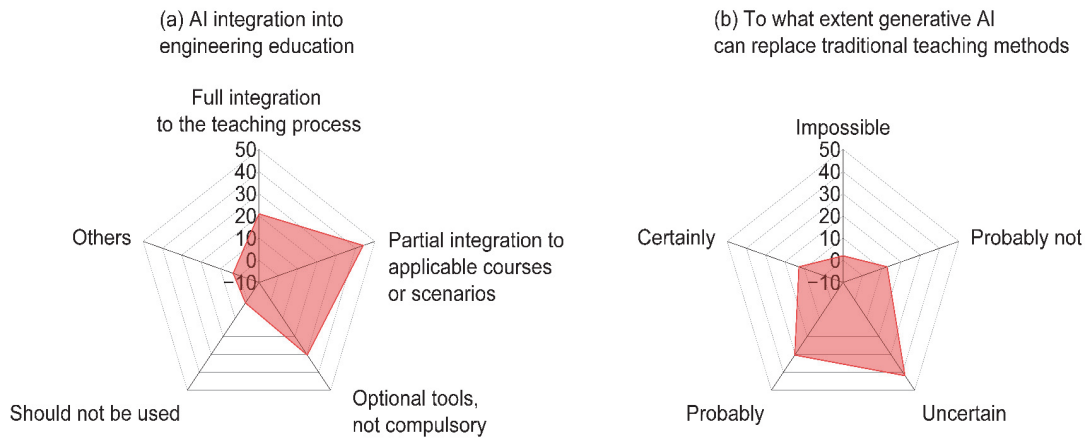

**Figure 6**: AI integration to engineering education from the perspective of Chinese engineering students: (a) Levels of integration, (b) Extent of replacement of traditional methods

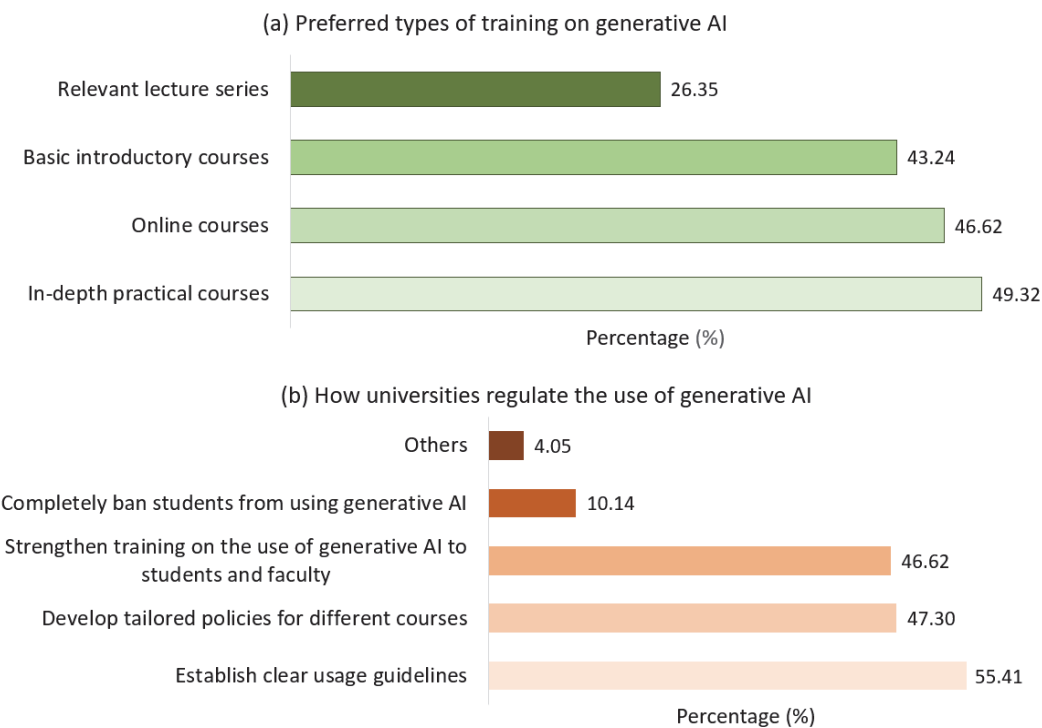

**Figure 7**. Institution supports on generative AI use: (a) Favored types of training by Chinese engineering students, (b) Types of regulations of AI use

Respondents also suggested several ways to regulate the use of generative AI, as shown in Figure 7(b). 55.41% believed that clear usage guidelines should be developed by universities, while 47.3% favored creating tailored policies based on the specific needs of each curriculum. 46.62% thought training on generative AI should be provided to both students and faculty. Interestingly, 10.14% supported a complete ban on generative AI use, highlight the concerns on the adoption of AI for academic purposes from a small group of students. Overall, the survey results suggest that most respondents prefer regulation and training over prohibition.

**5.4.2 Recommendations for educators and institutions**

As illustrated in Figure 8(a), most respondents were optimistic about the future of generative AI in education, with 29.05% and 39.86% describing its potential as "broad" or "very broad", respectively. However, 27.7% expressed a neutral stance, likely due to the current challenges and ethical concerns surrounding generative AI in engineering education. Only 3.38% held negative views, describing its prospects as "narrow" or "very narrow".

Respondents highlighted several areas for improvement in generative AI tools, as illustrated in Figure 8(b). Over 60% emphasized the need for higher accuracy in addressing discipline-specific problems, improved literature search capabilities, and integration with professional software, reflecting the higher academic and professional demands of engineering students. 54.73% expressed a desire for AI tools that can provide deeper insights into professional questions. There was also interest in enhancing AI's data-processing abilities, with 40.54% calling for improvements in this area.

Moving forward, schools and education policymakers should develop clear guidelines for the use of generative AI, design personalized integration plans tailored to different disciplines and curricula, and as well as provide comprehensive training programs covering technical operation, practical applications, and ethical issues.

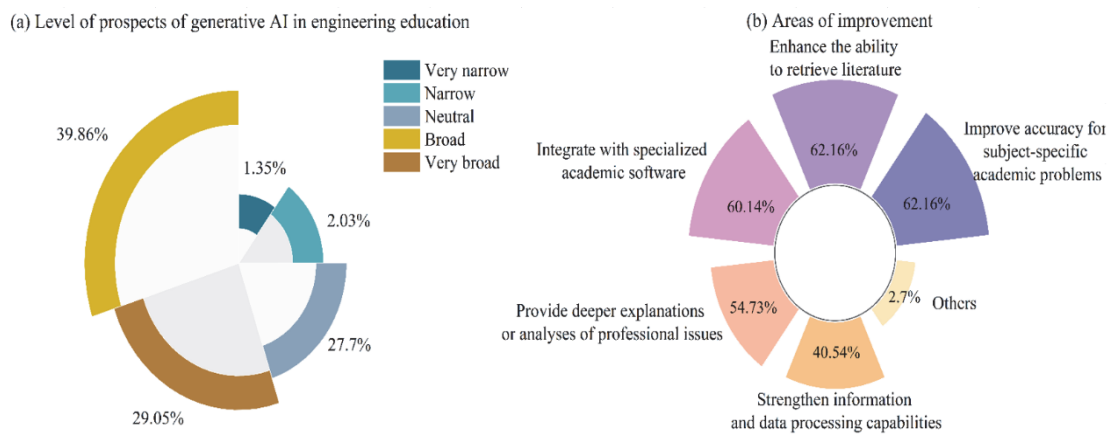

**Figure 8**. Prospects of generative AI in engineering education from Chinese engineering students' perspectives: (a) Level of prospects, (b) Areas of improvement

**6. Discussion**

**6.1 Implications for theory**

The use of generative AI among engineering students is expanding, with tools like ChatGPT leading the way due to their broad functionality and ease of use. From the perspective of the pedagogical theory TAM [49], generative AI is popular precisely because they provide powerful multimodal information generation and human language understanding capabilities that satisfy perceived usefulness, while highly user-friendly interaction design satisfies perceived ease of use, thereby reducing students' cognitive load in the technology adoption process, and improving acceptance of new technologies [49,50]. While postgraduate students used AI more frequently, the majority of students across all fields and all levels have relied on AI for tasks like academic writing, data analysis, and concept clarification. However, a smaller number of students remained hesitant or skeptical about the technology, suggesting a need for clearer guidance from educators on its proper use.

This study highlights that the generative AI has significantly enhanced learning efficiency and active learning of engineering students in China through the advantages of instant feedback and rapid content generation, and meanwhile improved the creativity and independent thinking of certain students. Reinforcement theory [40] points out that external rewards and punishments can effectively regulate learners' behavior [60]. External immediate feedback has a good effect in improving learning efficiency [55], and students' attention and creativity [61]. Unlike textbooks, which students generally assume to be accurate, AI-generated content can sometimes be incorrect, which encourages students to evaluate its accuracy when appropriately used. This can also cultivate independent thinking through the process of reflection and active construction of knowledge, which aligns well with the constructivist learning theory [45, 46] that emphasize active learning rather than passive reception of information. According to our survey, most surveyed students feel their independent thinking was improved or at least not affected.

When used irresponsibly, AI tools could lead to potential over-reliance and diminish students' ability to think independently, as suggested by a reasonable portion (17.56%) of respondents who expressed concerns about its decline. This behavior is driven by the immediate "reward" brought by AI feedback [40]. Every time a quick answer is obtained through AI, it positively reinforces students' tendency to avoid thinking and rely on external tools, thereby weakening their intrinsic motivation for independent thinking. Recent research has also confirmed this, reporting that this often occurs when individuals have difficulty assessing the reliability of the AI or how much to trust its recommendations [57, 58]. As such, while AI tools enhance learning in many ways, students must balance their use with active engagement in problem-solving and independent thought to fully benefit from their educational experience.

**6.2 Implications for practice**

Generative AI holds great promise for enhancing engineering education, but its application faces several challenges, including content accuracy, over-reliance on AI tools, and concerns about data privacy and ethics. Nearly two-thirds of surveyed students expressed their concern about the lack of accuracy of AI generated content, a critical issue in engineering where accuracy is essential. Inaccurate AI responses can mislead students, leading to errors in practical applications. Therefore, ensuring that AI-generated content is reliable and easily cross-verified is essential for its integration into engineering education.

Ethical concerns about generative AI in academic settings were highlighted, with 58.11% of respondents considering them "important" or "very important," though 40.54% viewed them as of "average importance." This underscores the need for regulation and ethics training to address these concerns. Data privacy also emerged as a significant issue, with 54.05% rating AI tools' privacy protection as "average" and only 37.84% expressing satisfaction, revealing a lack of confidence and room for improvement in this area. Addressing these challenges requires ongoing improvements in AI tool development, better user support, and more robust privacy protection. Furthermore, ethical guidelines should be established to ensure that AI tools are used responsibly and effectively in educational contexts.

Respondents identified key areas for improvement in generative AI tools, including better accuracy in discipline-specific problems, enhanced literature search, integration with professional software, and stronger data-processing capabilities. Engineering students also expressed a desire for AI to provide deeper insights into professional questions. To foster innovative growth in engineering education, schools should establish clear guidelines, create tailored integration plans, and offer comprehensive training on technical, practical, and ethical aspects of generative AI.

**6.3 Limitations and future research**

While this study provides valuable insights, it has several limitations. The reliance on self-reported data introduces potential biases, as students may exaggerate or downplay their use of AI tools [59]. Students may overestimate their learning with generative AI, perceiving progress despite little actual improvement [59]. This lack of accurate feedback can reduce motivation and engagement, ultimately hindering long-term learning depth. To reduce any potential subjective bias caused by self-reported data and the gap between perceived and actual learning outcomes, future studies can consider introducing multiple data sources for cross-validation, including but not limited to objective data such as students' test scores and homework scores, evaluation data from teachers or peers, and platform log data.

This study does not examine the long-term, evolving impact of generative AI on students' learning and performance. However, integrating AI technologies into students' learning activities and broadly higher education is inherently a gradual process, and short-term research may fail to capture their full influence on learning trajectories, core competency development, and career readiness. To gain a more comprehensive understanding of generative AI's role in engineering education, future research should adopt longitudinal approaches that track how AI integration affects skill development, academic outcomes, and professional preparedness over time.

The characteristics of participants' background may have important influences on the way they use generative AI tools. Regionally, students from areas with stronger educational resources are more open to adopting new technologies. The sample in this study may not be fully representative of all engineering students, particularly those from institutions with limited resources or slower technology adoption. In terms of disciplines, those in computer-related fields typically have greater technical skills and critical awareness, leading to more proactive AI use, while students in traditional engineering tend to use AI for specific tasks. Postgraduates are more inclined to apply AI tools to advanced academic work. Future research could explore these background factors in greater detail to better understand AI's impact on education. Additionally, future research can explore how AI can be tailored to the unique needs of different engineering disciplines. Furthermore, incorporating international survey data could expand the sample size and shed light on how cultural differences influence generative AI usage among engineering students.

Ethical and privacy concerns will continue to be a major area of study focus. Establishing robust ethical frameworks and policies is essential to protect students' rights and ensure that AI tools are used in a manner that aligns with educational values.

While the sampling strategy used in this study is appropriate and sufficient for its primary goal, the strategy could be enhanced through complementary methodologies to enable a more fine-grained and robust investigation. For greater representativeness, stratified random sampling could be employed to ensure proportional inclusion of students across different institution tiers and disciplines, though this would require more extensive coordination. Longitudinal study designs would allow researchers to capture the evolving nature of AI adoption over time. In addition, incorporating behavioral data analytics, such as AI platform usage metrics, could provide objective validation of self-reported behaviors and help reduce potential social desirability bias. A mixed-method approach, combining quantitative surveys with qualitative interviews, would offer deeper insights into usage patterns and the motivations behind them. While these alternatives demand more resources, they would significantly enhance the validity and generalizability of findings, supporting a more comprehensive and accurate understanding of how generative AI is shaping engineering education.

**7. Conclusion**

This study explores the use of generative AI among engineering students in China, examining its potential and challenges in engineering education through a questionnaire survey. The data and findings in this study offer valuable insights that can inform and support the developments of pedagogical theories.

The results indicate that a majority of students were already using generative AI in their academic and daily learning activities. The survey shows that generative AI has positively impacted learning efficiency, with 88.52% of respondents reporting improved productivity through faster content generation and feedback. 64.19% of participants experienced increased learning initiative, though 6.76% felt over-reliance negatively affected their motivation. Nearly 48% believed AI enhanced independent thinking, offering new perspectives, while a small minority felt it weakened their critical thinking. AI also boosted creativity for 58.78% of respondents, though 29.73% saw no change, and 11.48% felt it stifled creativity due to formulaic content. Nearly half of the respondents felt that using generative AI did not improve their academic performance, despite most acknowledging increased learning efficiency and greater engagement in active learning.

Generative AI in education presents several challenges, particularly in engineering, with students highlighting issues such as the inaccuracy of AI-generated content (62.16%), over-reliance on AI tools (39.86%), and technical difficulties with usability (20.27%). The inaccuracy of AI outputs, especially for complex tasks, undermines student confidence, while concerns about over-dependence may hinder independent problem-solving skills. Ethical concerns, privacy issues, and high costs were also identified as barriers to AI tool adoption. Additionally, while 42.57% of students found AI tools suitable for their needs, nearly 40% expressed doubts about AI's ability to handle specialized engineering problems, indicating a need for more adaptable and specialized AI solutions in the field.

The survey indicates strong optimism about the future of generative AI in education, with most respondents viewing its potential as broad, though 27.7% remained neutral due to challenges and ethical concerns. Regarding its integration into engineering education, 20.95% supported full integration, while 43.92% favored partial integration, with many believing generative AI courses should be compulsory.

For regulation, 55.41% favored clear usage guidelines, and 47.3% supported tailored policies based on curriculum needs, with 46.62% advocating for training for both students and faculty. Opinions on whether AI could replace traditional teaching methods were mixed, with 41.89% uncertain and 12.84% supporting the idea, though concerns over technical and ethical issues limited strong support. Respondents also emphasized the need for AI tools to improve accuracy, literature search capabilities, and integration with professional software. To move forward, educational institutions should focus on clear guidelines, customized integration plans, and comprehensive training programs covering technical, practical, and ethical aspects of generative AI.

## References

[1] Qadir J. Engineering education in the era of ChatGPT: Promise and pitfalls of generative AI for education. 2023 IEEE Global Engineering Education Conference (EDUCON). 2023;1–9. doi:10.1109/EDUCON54358.2023.10125121.

[2] Peres R, Schreier M, Schweidel D, Sorescu A. On ChatGPT and Beyond: How Generative Artificial Intelligence May Affect Research, Teaching, and Practice. Int J Res Mark. 2023;40(2):269–75. doi:10.1016/j.ijresmar.2023.03.001

[3] Fergus S, Botha M, Ostovar M. Evaluating Academic Answers Generated Using ChatGPT. J Chem Educ. 2023;100(4):1672–5. doi:10.1021/acs.jchemed.3c00087.

[4] Hwang G-J, Chang C-Y. A review of opportunities and challenges of chatbots in education. Interact Learn Environ. 2023;31(7):4099–112. doi:10.1080/10494820.2021.1952615.

[5] Miao F, Holmes W. Guidance for Generative AI In Education and Research. Paris, France: United Nations Educational, Scientific and Cultural Organization (UNESCO). 2023. doi:10.54675/EWZM9535.

[6] White J, Fu Q, Hays S, Sandborn M, Olea C, Gilbert H, et al. A prompt pattern catalog to enhance prompt engineering with ChatGPT. arXiv. 2023. doi:10.48550/arXiv.2302.11382.

[7] Bubeck S, Chandrasekaran V, Eldan R, Gehrke J, Horvitz E, Kamar E, et al. Sparks of artificial general intelligence: Early experiments with GPT-4. arXiv. 2023. doi:10.48550/arXiv.2303.12712.

[8] Khlaif ZN, Mousa A, Hattab MK, Itmazi J, Hassan AA, Sanmugam M, et al. The Potential and Concerns of Using AI in Scientific Research: ChatGPT Performance Evaluation. JMIR Med Educ. 2023;9:e47049. doi:10.2196/47049.

[9] Bak-Coleman J, Bergstrom CT, Jacquet J, Mickens J, Tufekci Z, Roberts T. Create an IPCC-like body to harness benefits and combat harms of digital tech. Nature. 2023;617(7961):462–4. doi:10.1038/d41586-023-01424-8.

[10] Wang WS, Lin CJ, Lee HY, Huang YM, Wu TT. Integrating feedback mechanisms and ChatGPT for VR-based experiential learning: Impacts on reflective thinking and AIoT physical hands-on tasks. Interact Learn Environ. 2024;1–18. doi:10.1080/10494820.2024.2375644.

[11] An S, Zhang S. Effects of ability grouping on students' collaborative problem solving patterns: Evidence from lag sequence analysis and epistemic network analysis. Think Skills Creat. 2024;51:101453. doi:10.1016/j.tsc.2023.101453.

[12] Deci, E. L., & Ryan, R. M. Intrinsic motivation and self-determination in human behavior. Plenum, 1985. ISBN 9780306420221.

[13] Wu JY, Liao CH, Tsai CC, Kwok OM. Using learning analytics with temporal modeling to uncover the interplay of before-class video viewing engagement, motivation, and performance in an active learning context. Computers and Education. 2024;212. doi:10.1016/j.compedu.2023.104975.

[14] UNESCO. Recomendación sobre la Ética de la Inteligencia Artificial. 2021. doi:10.54675/EWZM9535.

[15] Brown H, Lee K, Mireshghallah F, Shokri R, Tramèr F. What does it mean for a language model to preserve privacy? Paper presented at the ACM Conference on Fairness, Accountability, and Transparency. 2022:2280–92. doi:10.1145/3531146.3534642.

[16] Ellis AR, Slade E. A New Era of Learning: Considerations for ChatGPT as a Tool to Enhance Statistics and Data Science Education. J Stat Data Sci Educ. 2023;31(2):1–10. doi:10.1080/26939169.2023.2223609.

[17] Ausín T. ¿Por qué la ética para la Inteligencia Artificial? Lo viejo, lo nuevo y lo espurio. Sociol Tecnociencia. 2021;11:extra 2:1–16.

[18] Cooper G. Examining Science Education in ChatGPT: An Exploratory Study of Generative Artificial Intelligence. J Sci Educ Technol. 2023;32:444–52. doi:10.1007/s10956-023-10039-y.

[19] Wach K, Duong CD, Ejdys J, Kazlauskaitė R, Korzynski P, Mazurek G, et al. The dark side of generative artificial intelligence: A critical analysis of controversies and risks of ChatGPT. Entrep Bus Econ Rev. 2023;11(2):7–24. doi:10.15678/EBER.2023.110201

[20] Chen L, Chen P, Lin Z. Artificial Intelligence in Education: A Review. IEEE Access. 2020;8:75264–78. doi:10.1109/ACCESS.2020.2988510.

[21] Venkat Srinivasan. AI & learning: A preferred future. Computers and Education: Artificial Intelligence. 2022;3(100062-). doi:10.1016/j.caeai.2022.100062.

[22] Xu W, Ouyang F. The Application of AI Technologies in STEM Education: A Systematic Review from 2011 to 2021. International Journal of STEM Education. 2022;9. doi:10.1186/s40594-022-00377-5.

[23] Trist EL, Bamforth KW. Some Social and Psychological Consequences of the Longwall Method of Coal-Getting: An Examination of the Psychological Situation and Defences of a Work Group in Relation to the Social Structure and Technological Content of the Work System. Human Relations. 1951;4(1):3-38. doi:10.1177/001872675100400101.

[24] Mumford E. The story of socio-technical design: reflections on its successes, failures and potential. Information Systems Journal. 2006;16(4):317-342. doi:10.1111/j.1365-2575.2006.00221.x.

[25] Sarker, S., Chatterjee, S., Xiao, X., & Elbanna, A. (2019). The sociotechnical axis of cohesion for the IS discipline. MIS quarterly, 43(3), 695-A5. doi:10.25300/MISQ/2019/13747.

[26] Fui-Hoon Nah F, Zheng R, Cai J, Siau K, Chen L. Generative AI and ChatGPT: Applications, challenges, and AI-human collaboration. Journal of Information Technology Case & Application Research. 2023;25(3):277-304. doi:10.1080/15228053.2023.2233814

[27] Lv Z. Generative artificial intelligence in the metaverse era. Cognitive Robotics. 2023;3:208-217. doi:10.1016/j.cogr.2023.06.001.

[28] Boßelmann CM, Leu C, Lal D. Are AI language models such as ChatGPT ready to improve the care of individuals with epilepsy? Epilepsia (Series 4). 2023;64(5):1195-1199. doi:10.1111/epi.17570.

[29] Gu X, Chen C, Fang Y, Mahabir R, Fan L. CECA: An intelligent large-language-model-enabled method for accounting embodied carbon in buildings. Building and Environment. 2025;272:112694. doi.org/10.1016/j.buildenv.2025.112694

[30] Pantoja-Rosero BG, Oner D, Kozinski M, et al. TOPO-Loss for continuity-preserving crack detection using deep learning. Construction and Building Materials. 2022;344. doi:10.1016/j.conbuildmat.2022.128264. doi.org/10.1016/j.autcon.2025.106120

[31] Narazaki Y, Hoskere V, Yoshida K, Spencer BF, Fujino Y. Synthetic environments for vision-based structural condition assessment of Japanese high-speed railway viaducts. Mechanical Systems and Signal Processing. 2021;160. doi:10.1016/j.ymssp.2021.107850

[32] Kikalishvili S. Unlocking the potential of GPT-3 in education: Opportunities, limitations, and recommendations for effective integration. Interact Learn Environ. 2023;1–13. doi:10.1080/10494820.2023.2220401.

[33] Ahn SJ, Kim J, Kim J. The Bifold Triadic Relationships Framework: A Theoretical Primer for Advertising Research in the Metaverse. J Advert. 2022;51(5):592-607. doi:10.1080/00913367.2022.2111729.

[34] Museanu E. The Role of Ai in Learning Romanian as a Foreign Language for University Students. Journal of Information Systems & Operations Management. 2024;18(2):112-124. Accessed April 12, 2025. https://search.ebscohost.com/login.aspx?direct=true&db=bsu&AN=182781127&site=eds-live&scope=site

[35] Rui Sun, Xuefei "Nancy" Deng. Using Generative AI to Enhance Experiential Learning: An Exploratory Study of ChatGPT Use by University Students. Journal of Information Systems Education. 2025;36(1):53-64. doi:10.62273/ZLUM4022

[36] Huang YM, Wang WS, Lee HY, Lin CJ, Wu TT. Empowering virtual reality with feedback and reflection in hands-on learning: Effect of learning engagement and higher-order thinking. J Comput Assist Learn. 2024;40(4):1413–27. doi:10.1111/jcal.12959.

[37] Cheng SC, Hwang GJ, Chen CH. From reflective observation to active learning: A mobile experiential learning approach for environmental science education. Br J Educ Technol. 2019;50(5):2251–70. doi:10.1111/bjet.12845.

[38] Li H, Öchsner A, Hall W. Application of experiential learning to improve student engagement and experience in a mechanical engineering course. Eur J Eng Educ. 2019;44(3):283–93. doi:10.1080/03043797.2017.1402864.

[39] Gallent-Torres C, Zapata-González A, Ortego-Hernando JL. The impact of Generative Artificial Intelligence in higher education: a focus on ethics and academic integrity. RELIEVE. 2023;29(2):art. M5. doi:10.30827/relieve.v29i2.29134.

[40] Skinner BF. Selection by consequences. Science. 1981 Jul 31;213(4507):501-4. doi:10.1126/science.7244649.

[41] Escalante J, Pack A, Barrett A. AI-Generated Feedback on Writing: Insights into Efficacy and ENL Student Preference. International Journal of Educational Technology in Higher Education. 2023;20. doi:10.1186/s41239-023-00425-2

[42] Hao Q, Smith DH IV, Ding L, et al. Towards Understanding the Effective Design of Automated Formative Feedback for Programming Assignments. Computer Science Education. 2022;32(1):105-127. doi:10.1080/08993408.2020.1860408

[43] Banihashem SK, Kerman NT, Noroozi O, Moon J, Drachsler H. Feedback sources in essay writing: peer-generated or AI-generated feedback? International Journal of Educational Technology in Higher Education. 2024;21(1). doi:10.1186/s41239-024-00455-4

[44] Er E, Akçapınar G, Bayazıt A, Noroozi O, Banihashem SK. Assessing student perceptions and use of instructor versus AI-generated feedback. British Journal of Educational Technology. December 2024:1. doi:10.1111/bjet.13558

[45] Piaget J. PART I: Cognitive Development in Children: Piaget Development and Learning. Journal of Research in Science Teaching. 1964;2(3):176-186. doi:10.1002/tea.3660020306

[46] Piaget J. The Psychology of Intelligence. Routledge; 2001. Accessed April 11, 2025. ISBN 9780415254014

[47] Winkler R, Leimeister JM, Söllner M. Enhancing problem-solving skills with smart personal assistant technology. Computers and Education. 2021;165. doi:10.1016/j.compedu.2021.104148

[48] Banihashem SK, Farrokhnia M, Badali M, Noroozi O. The Impacts of Constructivist Learning Design and Learning Analytics on Students' Engagement and Self-Regulation. Innovations in Education and Teaching International. 2022;59(4):442-452. doi:10.1080/14703297.2021.1890634

[49] Davis FD. Perceived Usefulness, Perceived Ease of Use, and User Acceptance of Information Technology. MIS Quarterly. 1989;13(3):319-340. doi:10.2307/249008

[50] Li W, Zhang X, Li J, Yang X, Li D, Liu Y. An explanatory study of factors influencing engagement in AI education at the K-12 Level: an extension of the classic TAM model. Scientific Reports. 2024;14(1). doi:10.1038/s41598-024-64363-3

[51] Gardner H. Frames of Mind : The Theory of Multiple Intelligences. Basic Books; 2011. Accessed April 11, 2025. ISBN 9780465024339.

[52] Cronbach LJ. Coefficient alpha and the internal structure of tests. Psychometrika. 1951;16(3):297–334. doi:10.1007/BF02310555.

[53] Kaiser HF. An index of factorial simplicity. Psychometrika. 1974;39(1):31–6. doi:10.1007/BF02291575.

[54] Rodrigues de Souza R, Toebe M, Chuquel Mello A, Chertok Bittencourt K, Datsch Toebe IC. Optimizing Bartlett test: a grain yield analysis in soybean. Ciência Rural. 2023;53(6):1–5. doi:10.1590/0103-8478cr20220614.

[55] Fu, Q., Zheng, Y., Zhang, M. et al. Effects of different feedback strategies on academic achievements, learning motivations, and self-efficacy for novice programmers. Education Tech Research Dev 71, 1013–1032 (2023). https://doi.org/10.1007/s11423-023-10223-2

[56] Deci EL. Effects of Externally Mediated Rewards on Intrinsic Motivation. Journal of Personality and Social Psychology. 1971;18(1):105-115. Accessed April 11, 2025. https://search.ebscohost.com/login.aspx?direct=true&db=eric&AN=EJ035993&site=eds-live&scope=site

[57] Zhai C, Wibowo S, Li LD. The effects of over-reliance on AI dialogue systems on students' cognitive abilities: a systematic review. Smart Learning Environments. 2024;11(1). doi:10.1186/s40561-024-00316-7

[58] Klingbeil A, Grützner C, Schreck P. Trust and reliance on AI — An experimental study on the extent and costs of overreliance on AI. Computers in Human Behavior. 2024;160. doi:10.1016/j.chb.2024.108352

[59] Noroozi O, Banihashem SK, Alqassab M, Taghizadeh Kerman N, Panadero E. Does perception mean learning? Insights from an online peer feedback setting. Assessment and Evaluation in Higher Education. 2025;50(1):83-97-97. doi:10.1080/02602938.2024.2345669

[60] Ajjawi R, Boud D. Examining the nature and effects of feedback dialogue. Assessment & Evaluation in Higher Education. 2018;43(7):1106-1119. doi:10.1080/02602938.2018.1434128

[61] Qiming Rong, Qiu Lian, Tianran Tang. Research on the Influence of AI and VR Technology for Students' Concentration and Creativity. Frontiers in Psychology. 2022;13. doi:10.3389/fpsyg.2022.767689

**Appendix A**

| Variables | Scale-based questions |
|---|---|
| A. Impact of generative AI on individual learning capabilities | A1. How significant do you think generative AI tools are in improving your learning efficiency?<br>A2. How significant do you think generative AI tools are in influencing your ability to think independently?<br>A3. How significant do you feel generative AI has impacted your creativity?<br>A4. How significant do you feel the use of generative AI has altered your motivation for self-directed learning? |
| B. Perspectives on future applications of generative AI | B1. How optimistic do you feel the future development of generative AI in education over next five years?<br>B2. How broad do you think the potential applications of generative AI in engineering education?<br>B3. How significant do you believe generative AI will be for your future career growth?<br>B4. How optimistic are you about the use of generative AI in the field of engineering in the future?<br>B5. What is the level of your optimism regarding the innovative potential of generative AI in the field of engineering in the future?<br>B6. To what extent do you believe generative AI could replace traditional teaching methods? |
| C. Frequency of generative AI use across scenarios | C1. How often do you rely on generative AI tools as your primary source of ideas at the start of an assignment or research project?<br>C2. How often do you use generative AI tools to assist with literature reading tasks, such as summarizing or reviewing literature?<br>C.3 How often do you use generative AI tools to support your work or assignments in your professional courses or field of study?<br>C4. How often do you use generative AI tools to aid in writing reports? |
| D. Challenges and concerns about generative AI | D1. To what extent do you agree that generative AI is adequately adaptable and professional for your field?<br>D2. In what ways has the use of generative AI impacted your academic performance so far?<br>D3. How important do you consider ethical issues in the use of generative AI?<br>D4. How satisfied are you with the performance of generative AI in safeguarding data privacy?<br>D5. How often do you come across inaccurate information generated by AI tools during use? |
| E. Preference of using generative AI in complex problem-solving | E1. How likely do you intend to use generative AI tools for an initial search when exploring questions related to innovation and creativity?<br>E2. How likely do you intend to rely on generative AI tools as your primary solution when tackling interdisciplinary problems? |